\documentclass[12pt,a4paper]{article}
\usepackage{graphicx}
\usepackage{epsfig}
\usepackage{changebar}
\usepackage{lscape}
\usepackage[figuresright]{rotating}
\usepackage{amsmath}

\usepackage[warn]{mathtext}
\usepackage[T2A]{fontenc}
\usepackage[english]{babel}
\usepackage{psfrag}
\usepackage{amssymb}
\usepackage{xspace}

\begin{document}

\title{\Large \textbf Measurement of the energy dependence of $\sigma_{\rm tot}(\gamma p)$
with the ZEUS detector at HERA}

\author{{\slshape Amir Stern}  \\ {\small for the ZEUS Collaboration} \\[1ex]
\small{School of Physics and Astronomy, Tel Aviv University, 69978 Tel Aviv, Israel}}

\maketitle

\begin{abstract}
The energy dependence of the photon-proton total cross section,
$\sigma_{\rm tot}(\gamma p)$, was determined from $e^{+}p$ scattering data
collected with the ZEUS detector at HERA at three values of the
center-of-mass energy, $W$, of the $\gamma p$ system in the range
194$<W<$296 GeV. This is the first determination of the $W$ dependence
of $\sigma_{\rm tot}(\gamma p)$ from a single experiment at high $W$.
Parameterizing $\sigma_{\rm tot}(\gamma p) \propto W^{2 \epsilon}$,
$\epsilon$=0.111 $\pm$ 0.009 (stat.) $\pm$ 0.036 (syst.) was obtained.  
\end{abstract}

\section{Introduction}

Donnachie and Landshoff (DL)~\cite{Donnachie:1992ny} showed that the energy dependence of
all hadron-hadron total cross sections can be described by a simple
Regge motivated form,
\begin{equation}
\sigma_{\rm tot} = A \cdot (W^2)^{\epsilon} + B \cdot (W^2)^{-\eta} \, ,\label{totxsec}
\end{equation}
where $A$ and $B$ are process-dependent
constants, $W$ is the hadron-hadron center-of-mass energy,
and $\epsilon(=\alpha_{I\!\!P}(0)-1)$ and $\eta(=1-\alpha_{I\!\!R}(0))$ are
effective powers related to Pomeron and Reggeon exchange, respectively
($\alpha_{I\!\!P}(0)$ ($\alpha_{I\!\!R}(0)$) is the Pomeron (Reggeon) trajectory intercept).

The $\sigma_{\rm {tot}}(\gamma p)$ dependence on $W$ is particularly interesting
because of the nature of the photon, which is known to exhibit properties
of both a point-like particle (direct photon) and a hadron-like
state (resolved photon). At the $ep$ collider HERA,
$\sigma_{\rm {tot}}(\gamma p)$ can be extracted from $ep$ scattering at very low squared momentum
transferred at the electron vertex, $Q^2 \lesssim 10^{-3}$~GeV$^2$.

The measurements of the total $\gamma p$ cross section at
HERA for $W \simeq 200$~GeV~\cite{Derrick:1992kk,Ahmed:1992qc,Derrick:1994dt,Aid:1995bz,Chekanov:2001gw}
combined with measurements at low $W$ confirmed that the total photoproduction 
cross section has a $W$ dependence similar to that of
hadron-hadron reactions. However, the HERA measurements' systematic uncertainties were 
too large for a precise determination of the $W$ dependence of the
cross section. The original fits of DL gave $\epsilon=0.0808$ and no uncertainties
were determined. Cudell et al.~\cite{Cudell:1999tx} determined $\epsilon$ to be $0.093\pm0.003$.
However, only very few points were present for the highest center-of-mass energies.
The data were from different experiments and had a large spread.
Furthermore, the value of the Pomeron intercept comes out strongly correlated with that of the reggeon trajectories.
In another evaluation~\cite{Cudell:1996sh}, the authors give the range of 0.07--0.10 as acceptable values for $\epsilon$.

In the final months of operation, the HERA collider was run
with constant nominal positron energy, and switched to
two additional proton energies, 460 GeV and 575 GeV, lower than the nominal
value of 920 GeV. This opened up the possibility to determine precisely the
power of the $W$ dependence of $\sigma_{\rm {tot}}(\gamma p)$ from ZEUS data alone, in the range 194--296 GeV by 
measuring the ratios of cross sections, thus having many of the systematic uncertainties canceling out.

The difficulty in measuring total cross section, $\sigma$, in a collider
environment originates from the limited acceptance of collider detectors
for certain class of processes, in particular for elastic and diffractive scattering,
where the final state particles are likely to disappear down the beam-pipe.
The determination of the acceptance relies on Monte Carlo simulation of the physics and of
the detector. The simulation of the physics is subject to many uncertainties which then
impact on the systematic uncertainty of the cross section measurement.
For the energy dependence of $\sigma$, the impact of these uncertainties as well as of the
geometrical uncertainties can be minimized by studying the ratio $r$ of cross sections
probed at different $W$ values.

\noindent
Assuming $\sigma \sim W^{2\epsilon}$~\cite{Chekanov:2001gw},
\begin{equation}
r=\frac{\sigma(W_1)}{\sigma(W_2)}=\left(\frac{W_1}{W_2}\right)^{2\epsilon} \, .
\label{ratio}
\end{equation}
Experimentally,
\begin{equation}
\sigma = \frac{N}{A\cdot \cal{L}} \, ,
\label{xsec}
\end{equation}
where $A$, $\cal L$ and $N$ are the acceptance, luminosity and number of measured events,
respectively, and therefore
\begin{equation}
r=\frac{N_1}{N_2}\cdot \frac{A_2}{A_1} \cdot \frac{{\cal L}_{2}}{{\cal L}_{1}} \, ,
\label{acc}
\end{equation}
where the index 1(2) denotes measurements performed at $W_1$ ($W_2$). The acceptance for
$\gamma p$ events at HERA depends mainly on the detector infrastructure
in the positron (rear) direction. If the change in the $W$ value results
from changing the proton energy,
the acceptance is likely to remain the same, independently of $W$, and the ratio of
acceptances will drop out of formula~(\ref{acc}).

\section{Kinematics and cross section}

The photon-proton total cross section can be measured in the process
$e^+p \rightarrow e^+ \gamma p \rightarrow e^+ X$,
where the interacting photon is almost real.
The event kinematics may be described in terms of Lorentz-invariant
variables: the photon virtuality, $Q^2$,
the event inelasticity, $y$, and
the square of the photon-proton center-of-mass energy, $W$,
defined by
\begin{displaymath}
  Q^{2} = -q^{2} = -(k - k^{\prime})^2 \, , \; \; \; \; \; \; \; \;
  y = \frac{p\cdot q}{p\cdot k} \, , \; \; \; \; \; \; \; \;
  W^{2} = ( q + p )^{2} \, ,
\end{displaymath}
where $k$, $k^{\prime}$ and $p$ are the four-momenta of the
incoming positron, scattered positron and incident proton,
respectively, and $q = k - k^{\prime}$.
These variables can be expressed in terms of the experimentally
measured quantities
\begin{displaymath}
  Q^2 = Q^2_{\rm min} + 4 E_e E_e^{\prime} \sin^2\frac{\theta_e}{2} \, , \; \; 
  y = 1 - \frac{E_e^{\prime}}{E_e} \cos^2\frac{\theta_e}{2} 
  \simeq 1 - \frac{E_e^{\prime}}{E_e} \, , \; \;
  W \simeq 2 \sqrt{E_e E_p y} \, ,
\end{displaymath}
where $Q^{2}_{\rm min} = \frac{m^{2}_{e}y^{2}}{1 - y}$,
$E_{e}$, $E_{e}^{\prime}$ and $E_{p}$ are the energies of the
incoming positron, scattered positron and incident proton, respectively,
$\theta_e$ is the positron scattering angle with respect
to the initial positron direction and $m_e$ is the positron mass.
The scattered positron was detected in a positron tagger close to the beam
line, restricting  $\theta_e$ (and hence $Q^2$) to small values.
The photon virtuality ranged from the kinematic minimum,
$Q^{2}_{\rm min} \simeq 10^{-6}$~GeV$^2$, up to
$Q^2_{\rm max} \simeq 10^{-3}$~GeV$^2$,
determined by the acceptance of the positron tagger.

The $\gamma p$ cross-section can be determined from the measured number of $e^{+}p$ events using the following
formula:

\begin{equation}
\frac{d\sigma^{e^{+}p}(y)}{dy}=\frac{\alpha}{2\pi}\left[\frac{1+(1-y)^2}{y}ln\frac{Q^2_{\rm max}}
{Q^2_{\rm min}}-2\frac{(1-y)}{y}(1-\frac{Q^2_{\rm min}}{Q^2_{\rm max}})\right]\sigma^{\gamma p}_{\rm tot}(y)\, ,
\label{gammapxsec}
\end{equation}
\noindent
where $\alpha$ is the electromagnetic coupling constant. The term multiplying the $\gamma p$ cross-section in
known as the flux factor.

For each of the incident proton energies, $\sigma_{\rm tot}^{\gamma p}(y)$
has a small variation as a function of $y$ over the range of the measurement
and may be taken to be a constant, $\sigma_{\rm tot}^{\gamma p}$.
Thus, the flux may be integrated over the range of measurement to
give a total flux $F_{\gamma}$, which, when multiplied by the total
$\gamma p$ cross section gives $\sigma_{\rm tot}^{ep}$,
the ep cross section integrated over the measured range,
\begin{equation}
\sigma_{\rm tot}^{ep} = F_{\gamma} \cdot \sigma_{\rm tot}^{\gamma p} \; .
\label{eq-eptogp}
\end{equation}

\section{Experimental setup}

In the final days of HERA running, 27.5 GeV positrons were colliding with protons
of energy set at the nominal value of 920 GeV (high energy run, HER) and lowered
to 575 (medium energy run, MER) and 460 GeV (low energy run, LER).

\begin{figure}[htb]
\centerline{\includegraphics[width=0.9\textwidth]{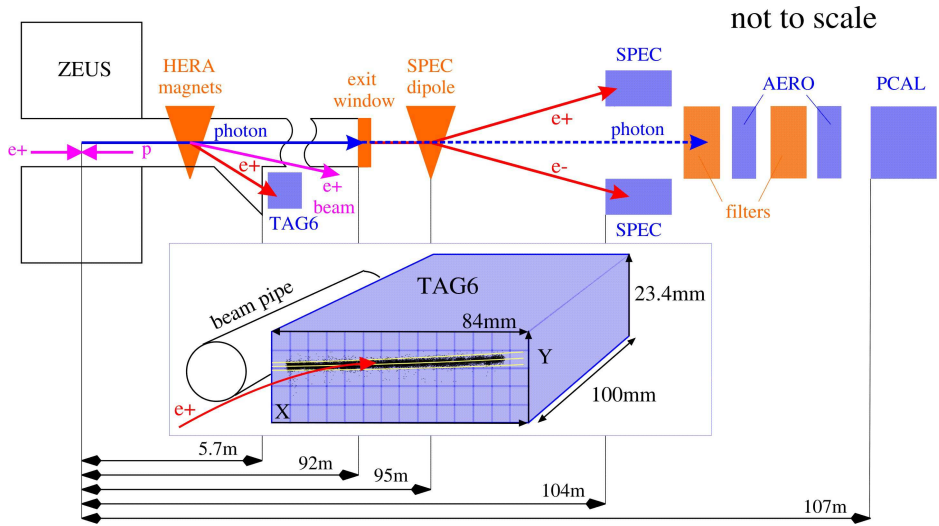}}
\caption{Schematic layout of the ZEUS detector, the six meter tagger and the components of the luminosity system
and their distance from the interaction point.}
\label{Fig:lumiline}
\end{figure}

The ZEUS detector components used in this analysis are the main
calorimeter (CAL) that consisted of three parts: the forward (FCAL),
the barrel (BCAL) and the rear (RCAL) calorimeters, the central
tracking detector (CTD), the microvertex detector (MVD),
the six meter tagger (TAG6) and the luminosity monitoring system.
The TAG6 was a spaghetti type calorimeter located 5.7~m from
the interaction point in the backward direction and was
used for tagging photoproduction events.
Scattered positrons were bent into it by the first
HERA dipole and quadrupole magnets after the interaction region,
with full acceptance for positrons with zero transverse momentum
in the approximate energy range 3.8--7.1 GeV with a $y$ range of 0.74--0.86.
The luminosity collected in the ZEUS detector was determined by two independent systems,
the photon calorimeter (PCAL) and the spectrometer (SPEC), that measured
the rate of the Bethe-Heitler (BH) process ($e^+ p \rightarrow e^+ \gamma p$).
The PCAL was shielded from primary synchrotron radiation by two carbon
filters, each approximately two radiation lengths deep.
Each filter was followed by an aerogel Cherenkov detector (AERO)
to measure the energy of showers starting in the filters.
The two components (PCAL and SPEC) enabled the measurement of the
luminosity in two independent ways with 1$\%$ relative uncertainty.
A schematic layout of the detector is shown in Fig.~\ref{Fig:lumiline}.

A special photoproduction trigger was implemented, requiring a low-angle scattered positron
candidate detected in the TAG6 and some activity in RCAL.
To reduce the background from events with energy in RCAL and a TAG6 hit caused by a random coincidence with a
BH event in the same HERA bunch, the energy in the PCAL,
$E_{\rm PCAL}$, was restricted to $E_{\rm PCAL} \lesssim$ 14 GeV.

The {\sc pythia} 6.416~\cite{pythia} generator, coupled to the {\sc heracles} 4.6~\cite{heracles,heracles46}
program (to simulate electromagnetic radiative effects), was used to simulate the photoproduction processes
in the proper weight to describe the CAL energy distributions in the total-cross-section data.

The acceptance for $\gamma p$ events is determined by the acceptance of the TAG6
and that of the main detector, which are independent.
The above mentioned simulation was used to calculate the acceptance of the RCAL for photoproduction
processes.

\section{Event selection and data analysis}

Clean positron hits in the TAG6 were selected by requiring
that the highest-energy cell was not at the edge of the detector.
Showers from inactive material in front of the tagger were rejected
by a cut on the energy sharing among towers surrounding the
tower with highest energy.
The position of the positron was reconstructed
by a neural network trained on an MC simulation
of the TAG6~\cite{thesis:gueta:2010}.
The neural-network method was also used to correct the energy
of the positrons for a small number of noisy cells, which were excluded.
Events from the BH process, selected
by requiring a positron in the TAG6 in coincidence with
a photon in the SPEC, were used to calibrate the TAG6 with positrons
with very small transverse momentum.
The energy, $E$, was determined as a function of the horizontal
position, $X$, and the
correlation between $X$ and the vertical position, $Y$, was also measured.
Cuts were placed on $E(X)$ and $Y(X)$ for the photoproduction events to reject
positrons with transverse momentum $p_T \gtrsim 10$ MeV,
off-momentum beam positrons,
and background from beam-gas interactions~\cite{thesis:gueta:2010}.
The $(X,Y)$ distribution of positrons from a sample of BH events
from the MER,
and the $Y(X)$ cuts, are shown in the inset in Fig.~\ref{Fig:lumiline}.

In RCAL, the towers immediately horizontally adjacent to the
beam-pipe hole had a large rate from off-momentum beam positrons
and debris from beam-gas interactions which satisfied the trigger conditions.
In events in which the RCAL cell with highest energy was in one
of these towers, the fraction of total RCAL energy, $E_{\rm RCAL}$, in
that tower was required to be below an $E_{\rm RCAL}$-dependent
threshold~\cite{thesis:gueta:2010}.
This eliminated most of the background and resulted
in only about 2.9\% loss of signal events.

Background from positron beam-gas interactions passing the trigger
requirement was determined from non-colliding HERA positron bunches.
This sample was subtracted statistically from the colliding
HERA bunches by the ratio of currents of $ep$ bunches to $e$-only
bunches.

Photoproduction events associated with the TAG6 hit
could have a random coincidence with
an event in the same HERA bunch from the BH process,
with the BH photon depositing more than 14 GeV in
the PCAL and therefore vetoing the event.
To account for this loss, accepted events were weighted
by a factor determined from the
rate of overlaps at the time the event was accepted.
The fraction of overlaps is proportional to the instantaneous
luminosity, which was higher during the HER relative to the LER and MER.
The correction for this effect was $\approx +2.6$\% for
the HER and  $\approx +1.2$\% for the LER and MER data samples.

Another background came from photoproduction events outside
the $W$ range of the TAG6 but satisfying the RCAL trigger,
with a random coincidence from BH hitting the TAG6.
The photon from the BH event may not have been vetoed
by the $E_{\rm PCAL} \lesssim 14$ GeV requirement
due to the limited acceptance and resolution of the PCAL. Such overlaps were studied using the distribution of
the energy of the PCAL+AERO; this offered greatly
improved photon energy resolution over the PCAL alone.
In addition to the BH events which produced a TAG6 hit,
this spectrum also contained photoproduction events
associated with the TAG6 hit overlapping in the same HERA bunch
with a photon from a random BH event whose positron did not hit the TAG6.
The number of overlaps seen in the PCAL, corrected for the PCAL acceptance,
was the number of BH overlaps to subtract from the selected photoproduction sample.

\section{Results}

The total photon-proton cross section for one proton energy is given by
\begin{equation}
\sigma_{\rm tot}^{\gamma p} = \frac{N}{{A_{\rm RCAL}} \cdot {F_{\gamma}^{\rm TAG6}} \cdot {\cal{L}}} \, ,
\end{equation}
where $N$ is the measured number of events, $\cal{L}$ is the
integrated luminosity, $F_{\gamma}^{\rm TAG6}$ is the fraction of the photon flux
tagged by the TAG6, and $A_{\rm RCAL}$ is the acceptance
of the hadronic final state for tagged events.

The acceptance of the detector for all three energy settings was found to be equal within errors and
thus their ratio cancels. The data taken correspond to a luminosity of 567 nb$^{-1}$ in the
HER, 949 nb$^{-1}$ in the MER and 912 nb$^{-1}$ in the LER.

\begin{figure}[htb]
\centerline{\includegraphics[width=0.65\textwidth]{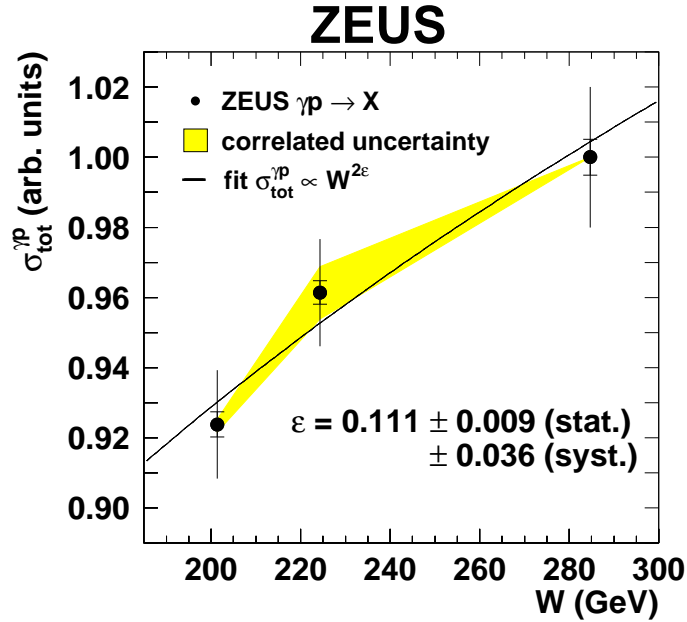}}
\caption{The $W$ dependence of the total photon-proton cross section,
normalized to the value for the HER. The inner error bars show the
statistical uncertainties of the total-cross-section data; the outer error
bars show those uncertainties and all uncorrelated systematic
uncertainties added in quadrature. The shaded band shows the effect
of the correlated systematic uncertainties. The curve shows the
fit to the form $\sigma_{\rm tot}(\gamma p) \propto W^{2 \epsilon}$.}
\label{Fig:wdep}
\end{figure}

In Fig.~\ref{Fig:wdep} the measured relative values of $\sigma_{\rm tot}^{\gamma p}$
are shown as a function of $W$, where the cross section for HER is normalized to unity.
A fit of the form $W^{2\epsilon}$ was performed to the relative cross sections
using only the statistical uncertainties, and separately with all the uncorrelated
systematic uncertainties added in quadrature. The correlated shifts were then applied
to the data and the fit repeated; the change in $\epsilon$ was negligible.
The result for the logarithmic derivative in $W^2$ of the energy dependence is
\begin{equation*}
\epsilon = 0.111 \pm 0.009 \, {\rm (stat.)} \pm 0.036 \, {\rm (syst.)} \, .
\end{equation*}

\noindent
This result is consistent with earlier determinations of $\epsilon$, however has the advantage of being obtained
from a single experiment.

In the picture in which the photoproduction cross section is
$\propto \ln^2(W^2)$ as required by the Froissart bound~\cite{Froissart},
$\epsilon \approx 0.11$ is expected, in agreement with the present measurement.
The interpretation of this result in terms of the Pomeron intercept is
subject to assumptions on the Reggeon contribution in the
relevant $W$ range. The most recent analysis of all hadronic cross sections
using a fit taking into account Pomeron and Reggeon
terms~\cite{Cudell:2001pn} yielded a Pomeron intercept of 0.0959 $\pm$ 0.0021. This is
in agreement with the result presented here.

\begin{footnotesize}

\end{footnotesize}


\end{document}